%% file: main.tex
\documentclass[fleqn,10pt]{wlscirep}
\usepackage[utf8]{inputenc}
\usepackage[T1]{fontenc}
\usepackage{color,soul}
\usepackage{amsmath}
\usepackage{hyperref}
\usepackage{setspace}
\hypersetup{
	colorlinks=true,
	linkcolor=blue,    
	urlcolor=cyan,
}

\graphicspath{{Figures/}}

\title{Enhanced hyperspectral tomography for bioimaging by spatiospectral reconstruction}
\author[1,*]{Ryan Warr} 
\author[1,2]{Evelina Ametova}
\author[1]{Robert J. Cernik}
\author[3]{Gemma Fardell}
\author[4]{Stephan Handschuh}
\author[5,6]{Jakob S. J\o{}rgensen}
\author[1,3]{Evangelos Papoutsellis}
\author[3]{Edoardo Pasca}
\author[1]{Philip J. Withers}
\affil[1]{Henry Royce Institute, Department of Materials, The University of Manchester, Manchester, M13 9PL, UK}
\affil[2]{Laboratory for Applications of Synchrotron Radiation, Karlsruhe Institute of Technology, Kaiserstr. 12, D-76131, Karlsruhe, Germany}
\affil[3]{Scientific Computing Department, Science
Technology Facilities Council, UK Research and
Innovation, Rutherford Appleton Laboratory, Chilton, Didcot, OX11 0QX, UK}
\affil[4]{VetCore Facility for Research, University of Veterinary Medicine Vienna, Vienna, Austria}
\affil[5]{Department of Applied Mathematics and Computer
Science, Technical University of Denmark, Denmark}
\affil[6]{Department of Mathematics, The University of Manchester, Manchester, M13 9PL, UK}

\affil[*]{Correspondence should be addressed to: \url{ryan.warr@postgrad.manchester.ac.uk}}

\keywords{hyperspectral imaging, iterative reconstruction, analytic reconstruction, regularisation, k-edge subtraction, Anolis species.}

\begin{abstract}
Here we apply hyperspectral bright field imaging to collect computed tomographic images with excellent energy resolution (800 eV), applying it for the first time to map the distribution of stain in a fixed biological sample through its characteristic K-edge. Conventionally, because the photons detected at each pixel are distributed across as many as 200 energy channels, energy-selective images are characterised by low count-rates and poor signal-to-noise ratio. This means high X-ray exposures, long scan times and high doses are required to image unique spectral markers. Here, we achieve high quality energy-dispersive tomograms from low dose, noisy datasets using a dedicated iterative reconstruction algorithm. This exploits the spatial smoothness and inter-channel structural correlation in the spectral domain using two carefully chosen regularisation terms. For a multi-phase phantom, a reduction in scan time of 36 times is demonstrated. Spectral analysis methods including K-edge subtraction and absorption step-size fitting are evaluated for an \emph{ex vivo}, single (iodine)-stained biological sample, where low chemical concentration and inhomogeneous distribution can affect soft tissue segmentation and visualisation. The reconstruction algorithms are available through the open-source Core Imaging Library. Taken together, these tools offer new capabilities for visualisation and elemental mapping, with promising applications for multiply-stained biological specimens.
\end{abstract}

\begin{document}

\flushbottom
\maketitle

\thispagestyle{empty}

\input{introduction.tex}

\input{results.tex}

\input{discussion.tex}

\input{conclusion.tex}

\input{methods.tex}

\bibliography{References}
\section*{Acknowledgements}
We acknowledge the following EPSRC grants for funding that have enabled this project: "A Reconstruction Toolkit for Multichannel CT" (EP/P02226X/1) and "CCPi: Collaborative Computational Project in Tomographic Imaging" (EP/M022498/1 and EP/T026677/1). PJW and RW acknowledge support from the European Research Council grant No. 695638 CORREL-CT. JSJ was partially supported by The Villum Foundation (grant no. 25893). EA was partially funded by the Federal Ministry of Education and Research (BMBF) and the Baden-Württemberg Ministry of Science as part of the Excellence Strategy of the German Federal and State Governments. The Manchester (Henry Moseley) X-ray Imaging Facility was funded in part by the EPSRC (grants EP/F007906/1, EP/F001452/1 and EP/M010619/1). This work makes use of computational support by CoSeC, the Computational Science Centre for Research Communities, through CCPi.  

\section*{Author contributions statement}

R.W., S.H., P.J.W. and R.J.C. conceived the experiments, S.H. and R.W. produced the samples for experimentation. R.W. conducted the experiments with support from J.S.J., E.A. and E.Pap. Software was developed by E.A., G.F., J.S.J., E.Pap. and E.Pas. R.W. wrote the manuscript, with critical reviews and support from all authors.

\section*{Additional information}

\textbf{Data Availability:} The phantom sample datasets generated and analysed during the current study are available from \url{https://doi.org/10.5281/zenodo.4354816}, while the biological sample dataset is available from \url{https://doi.org/10.5281/zenodo.4352944} \\
\textbf{Competing interests:} The authors declare no competing interest.

\end{document}

%% file: introduction.tex
\section*{Introduction} \label{sec:introduction}
With the increasing development in X-ray detector technology, the interest in energy-selective tomography has grown in recent years, particularly for medical imaging. In conventional X-ray absorption computed tomography (CT), each detector pixel records the total number of detected photons, irrespective of their energy, building up a single radiograph for each projection angle. Contrast is hence solely provided by differences in attenuation based on a sample’s local material composition. Issues may then arise in post-processing segmentation, particularly for phases or structures of similar electron density, such as different types of soft tissue. In addition, due to the use of polychromatic radiation in laboratory X-ray imaging, beam-hardening artefacts are common, as the non-linear nature of attenuation as a function of energy is ignored without energy discrimination \cite{Davis2008}. The introduction of spectroscopic detectors have enabled an additional dimension of information to be acquired, by measuring both the energy and position of each incident photon. Given that every element has a unique attenuation profile, these may be used as spectral ‘fingerprints’ in energy-sensitive CT imaging, such that full 3D elemental mapping may be performed from a single CT scan. 

Recent spectral detectors take two main forms, differing mainly in their measurement processes, and number of energy ‘channels’ used for photon binning. Multispectral detectors use a set of 4-8 threshold (reference) channels, to which each incident photon is allocated based on the electrical signal generated upon detection. Such detectors have found use in the field of medical imaging for soft tissue differentiation  \cite{Prebble2018,Aamir2014,Anderson2010,Anderson2014,Roeder2017}. While the low channel number enables high count-rates similar to conventional CT imaging  \cite{Ballabriga2016}, multispectral imaging provides coarse energy resolution (\textasciitilde 5-10 keV), thus for spectrally similar species, the threshold positions require specific pre-selection, as well as \emph{a priori} knowledge of sample composition. In contrast, hyperspectral detectors offer very fine energy resolution (< 1 keV) by storing photons into hundreds of narrow energy channels. The result is that, for every pixel, we acquire a 'pixel spectrum', containing a full absorption profile, for studying changes in attenuation as a function of energy. A key factor for chemical fingerprinting is the presence of absorption edges, observed as sharp discontinuities at the energies equivalent to the binding energies of the core-electron states (e.g. K-edges). Absorption edges act as characteristic markers for chemical identification, with many of these edges falling within the hard X-ray range (> 10 keV) used for imaging. One method that utilises the absorption edges is dual-energy CT (DECT). Following two sequential scans taken at different energies, DECT decomposes materials by virtue of their differing attenuation as a function of energy  \cite{Johnson2012}. In biological and medical imaging, contrast-enhanced investigations in DECT use highly attenuating chemical tracers which preferentially bind to specific tissue structures. Numerous studies have previously reviewed the advantages of various contrast agents (e.g. I$_{2}$, PMA/PTA) in both DECT and conventional CT for \emph{in vivo} and \emph{ex vivo} biological imaging \cite{Descamps2014,Handschuh2017,Metscher2009,Gignac2016,Badea2012}. A common limitation of DECT, however, is its time-consuming nature and the dose considerations required in some cases \cite{Achenbach2008,Schenzle2010}. Further, the technique struggles in the differentiation of materials with spectrally similar characteristics, and therefore cannot unequivocally distinguish all materials. Hyperspectral imaging has, on the other hand, previously been demonstrated for the separation of closely-spaced K-edges, for example in the non-destructive evaluation of mineralised ore samples \cite{Egan2015}.

The capability of identifying spectral markers from the detailed absorption profile of each detector pixel comes at the cost of increased signal processing during data acquisition. Because the detected photons are distributed across hundreds of energy channels, the count-rate is limited to levels far below that of conventional CT \cite{Ballabriga2016}. Any individual energy bin is therefore subject to low signal-to-noise ratio (SNR) for each measured pixel spectrum. The issue of low SNR can create problems in chemical identification, particularly in the case of low concentration, where weaker signals may be hidden amongst the surrounding noise. Increased exposure time is one solution for improving SNR, however this is not always possible. Alternatively, the choice of reconstruction algorithm may be optimised for improved image quality. A conventional cone-beam reconstruction algorithm, such as Feldkamp-Davis-Kress (FDK - the 3D form of filtered back-projection for cone-beam imaging \cite{Feldkamp1984}) often fails to accurately reconstruct features for low count or undersampled data \cite{Buzug}. An advantage of hyperspectral imaging is the strong structural correlations between channels. Due to the fine energy resolution of the detectors, narrow channels result in a high degree of similarity between neighbouring energy bins. It is therefore possible to exploit the ’channel-wise’ nature of the dataset as part of the reconstruction process, by employing dedicated algorithms. In the case of spectral imaging, this correlative nature may be used to, for example, provide both noise reduction and feature preservation in the spatial or spectral domains \cite{Egan2017a}. A number of methods have previously been evaluated for their effectiveness in spectral image reconstruction, including both undersampled and noisy datasets \cite{Kazantsev2018,Rigie2015,Xu2014}. However, the availability of advanced, spectral reconstruction methods is still currently limited and inconsistent across the field. A set of modules have recently been developed in the form of open source software - the Core Imaging Library (CIL: \url{http://www.ccpi.ac.uk/CIL}) - to aid the complete workflow of CT datasets, including for 4D acquisition modalities like spectral imaging. Each method has been optimised for fast, simple use across a range of geometries and techniques. A detailed description of CIL may be found elsewhere, for both the overall software \cite{Jorgensen2020} and for its hyperspectral capabilities \cite{Papoutsellis2020}. 

In this paper, we examine the capability of advanced spectral reconstruction methods in providing high quality results from undersampled, noisy 4D datasets, to enable elemental mapping. Using the HEXITEC hyperspectral detector \cite{Seller2011} in a lab-based setting, we set out two main aims. Firstly, we assess the impact of implementing a dedicated, iterative reconstruction algorithm, compared to the conventional FDK method, using a physical, multi-phase phantom. By directly comparing reconstructions following a significant reduction in scan time, we evaluate the extent of feature restoration both in the spatial and spectral domains. Secondly, we evaluate reconstruction quality on a biological sample by means of different spectral analyses, including segmentation by K-edge subtraction, and measurement of relative chemical concentration. Once more, a correlated reconstruction algorithm is applied, and used to highlight the vulnerabilities of the FDK method for noisy datasets. A single-stained lizard head was considered for mapping of tissue structures, and compared to equivalent results obtained following DECT. The work evaluates the advantages of hyperspectral imaging, compared to DECT and multispectral CT, where poorer energy resolution limits further analysis. This study also opens the way for further investigations into simultaneous staining of multiple contrast agents, where little work has previously been performed outside of DECT \cite{Handschuh2017}, or phantom studies for multispectral imaging \cite{Mahdieh2016,Panta2018,Anderson2010}.

%% file: results.tex
\section*{Results} \label{sec:results}
\begin{figure}[ht!]
	\centering
	\includegraphics[scale=0.55]{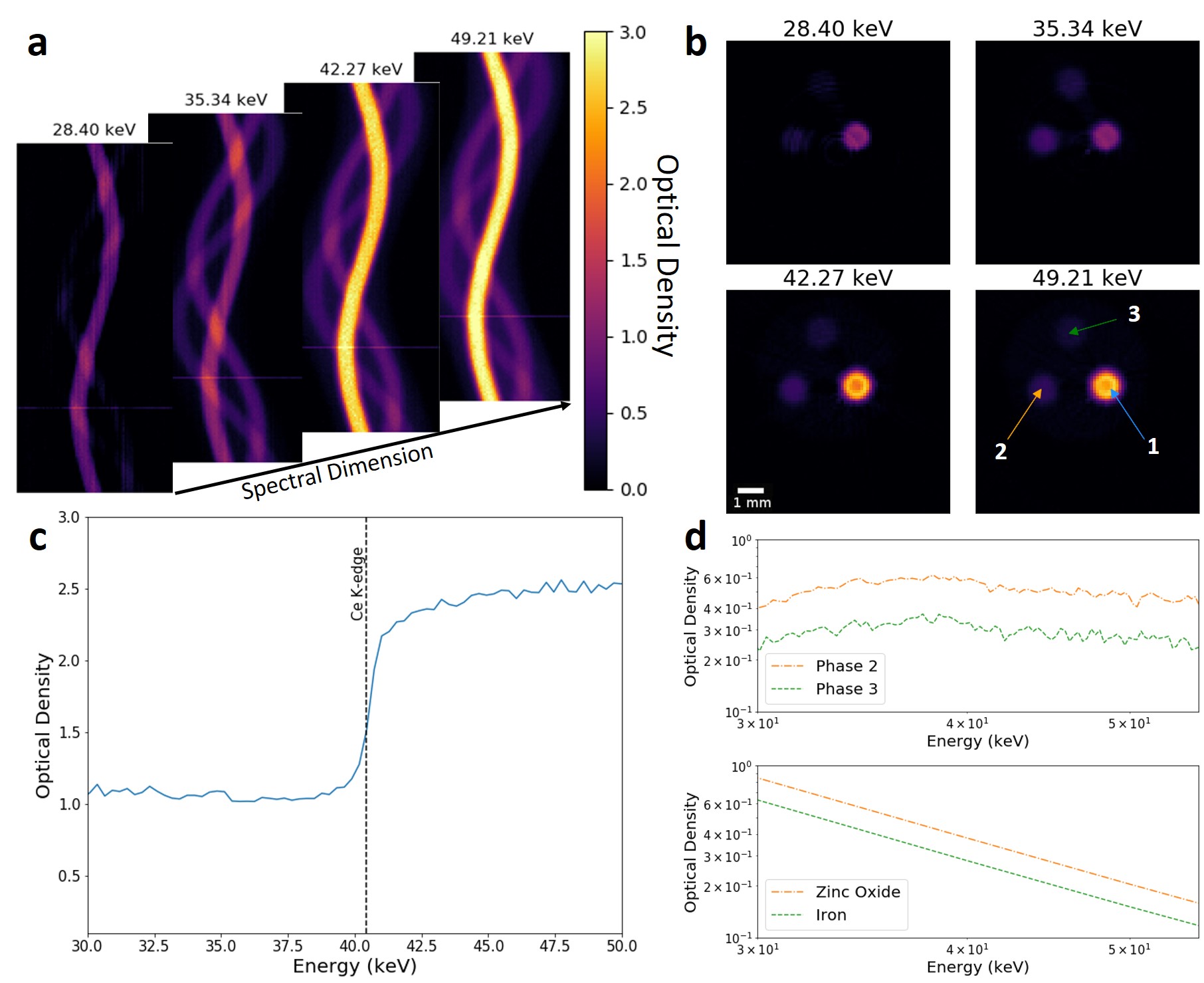}
	\caption{\textbf{Optical density variation in the spectral dimension for a multi-phase powder phantom}. \textbf{(a)} Set of four sinograms taken between channels 100-175 at equal spacing of 25 channels (corresponding energies are shown) for Scan A. A single discontinuity in the sinograms appears due to an interrupted scan. \textbf{(b)} The corresponding FDK reconstructions for each energy channel of the sinograms in \textbf{(a)}. Three distinct regions are observed, corresponding to the three metal-based powders. The colour scale measures the dimensionless unit OD, and is consistent across both images. Three voxels are highlighted (marked by numbered arrows for each respective powder phase). \textbf{(c)} Single voxel spectra of powder phase 1. A line signifying the theoretical position of the cerium K-edge is overlaid for comparison. \textbf{(d)} Comparison of measured absorption spectra (top) for voxels 2 and 3 located in the zinc oxide and iron phases respectively, and the theoretical values (bottom) over the same spectral range.}
	\label{fig:powder_recon_sino_spec}
\end{figure}

Firstly we examine a simple aluminium phantom test sample, filled with 3 different powders (CeO$_{2}$, ZnO and Fe). Comprising different elements, each powder has different characteristic absorption properties for analysis. The phantom was imaged using a micro-focus source at 60 kV with beam power 6 W. Two scans were acquired using different exposure times, with an effective voxel size of 98 $\mu$m. The full dimensions of the sample were captured within the 80 $\times$ 80 detector field of view (FOV). The number of channels used in hyperspectral imaging are determined by two system characteristics - the tube voltage and the energy-channel calibration. For the 60 kV tube voltage applied, the full spectral profile was segmented across 200 energy channels for each pixel. Given we have \emph{a priori} knowledge of sample composition and, thus, absorption edge position in this case, a reduced subset of channels were selected for both datasets (corresponding to channel numbers 100 - 200, or energy range 28 - 56 keV), in order to increase the speed of reconstruction. To evaluate the capabilities of the iterative method in handling low count and under-sampled datasets, a comparison was made between two acquisition schemes: Scan A was taken with 180 projections and 180 s exposure time, and Scan B was taken with 30 projections and 30 s exposure time. Through prior testing, it was found that the conditions used for Scan A produced sufficiently low noise and strong feature definition, such that it may be treated as a 'ground truth', optimum state, when using conventional channel-wise FDK reconstruction. All results obtained under the conditions of Scan B were then compared directly to this reference case. Given the nature of spectral datasets, each channel may be analysed on an individual basis, akin to a set of multiple monochromatic scans. Figure \ref{fig:powder_recon_sino_spec}a illustrates a set of sinograms for Scan A, corresponding to four equally spaced energy channels, combined with the corresponding central slice FDK reconstructions for each energy. An important factor to note is that each voxel value, $u$ is measured in terms of the dimensionless optical density (OD). That is, by rearrangement of the Beer-Lambert law, $\mu u = \ln(I_{0}/I)$, changes in X-ray attenuation may be measured as a function of energy, highlighting key spectral features such as absorption edges. We avoid use of the absolute value of the attenuation coefficient, $\mu$, as this is unreliable in hyperspectral imaging, mostly due to the effect of 'charge-sharing' between pixels, which can result in erroneous photon energies during scanning. For instances of lower X-ray flux, this may be corrected for, as detailed elsewhere \cite{Egan2015}.

Through examination of the monochromatic reconstructions in Fig. \ref{fig:powder_recon_sino_spec}b (covering the energy range 28 - 49 keV), we observe a significant rise in OD for one such phase, indicative of a spectrally significant elemental marker. By analysing individual voxel absorption spectra, chemical insight on sample composition can be obtained by examining changes in OD as a function of energy. Three voxels were selected in Fig. \ref{fig:powder_recon_sino_spec}b. The spectra shown in Fig. \ref{fig:powder_recon_sino_spec}c corresponds to phase 1 where, as expected, an absorption edge is observed at 40.4 keV. Owing to the high energy resolution of the system, one may conclusively confirm that such an edge corresponds to that of cerium (40.443 keV). For the two remaining material phases in Fig. \ref{fig:powder_recon_sino_spec}d, little variation in OD is observed. In addition, no absorption edges belonging to these materials (ZnO, Fe) were seen, as they fall below the sensitivity of the hyperspectral system (< 10 keV). Other methods may be employed to segment such materials, for instance through measuring the relative change in OD as a function of energy. To highlight this point, Fig. \ref{fig:powder_recon_sino_spec}d includes a plot of the theoretical optical density values for both ZnO and Fe, across the same energy range. The similarity of measured phase values to the predicted data, as well as the difference in magnitude between them, offers a method of material identification and segmentation, absent of absorption edges, as has been shown elsewhere \cite{Egan2015}. Such analysis is not the focus of this paper. Unsurprisingly, the conditions of Scan A provide high quality reconstructions, with sharp feature definition, and little noise fluctuations across the full spectral range. Next we will explore differences when reconstructing the low count and undersampled dataset of Scan B. 

As previously discussed, analytic reconstruction methods, such as FDK, often fail to adequately reconstruct features for low SNR data. Here we explore the extent to which noisy, few projection data can be reconstructed, through the use of a spatiospectral reconstruction algorithm. The reconstruction problem is typically formulated as a combination of a data fitting term between measured and reconstructed data, and one or more regularisation terms which encode desirable image characteristics. The problem is then solved using an iterative optimisation algorithm. CIL provides a number of building blocks to formulate the optimisation problems, and solvers to find a numerical solution. In this case we used a combination of two regularisation terms, known as the Total Variation (TV) and Total Generalised Variation (TGV), implemented along the spatial and spectral dimensions respectively. Further details of the regularisation terms, and how they were applied to the dataset, can be found in the methods section. The joint regularisation method, referred to here onwards as TV-TGV, is a novel method, chosen based on the prior knowledge that we expect noisy images, combined with the presence of an absorption edge. The CIL software enables such a method to be constructed and applied, and its advantages over other state-of-the-art methods has been demonstrated elsewhere in the case of Bragg-edge neutron tomography \cite{Ametova2020}.

For the reconstruction of Scan B, TV-TGV was applied simultaneously across the full set of energy channels. By doing so, we exploit the correlations between neighbouring channels, compared to FDK, which is applied channel-by-channel and therefore cannot make use of such structural similarities. As described in Equation \ref{eq:it_recon} (see methods section), three regularisation parameters ($\alpha$ for TV, $\beta_{1,2}$ for TGV) were optimised. Final parameters were determined to be 0.002, 0.18 and 0.25 for $\alpha$, $\beta_{1}$ and $\beta_{2}$ respectively. A range of iteration numbers were tested, with no discernible improvements in image quality observed beyond 1000 iterations. Final reconstruction of the full volume was achieved with a runtime of 25 minutes for 1000 iterations of the algorithm.

\begin{figure}[ht!]
	\centering
	\includegraphics[scale=0.65]{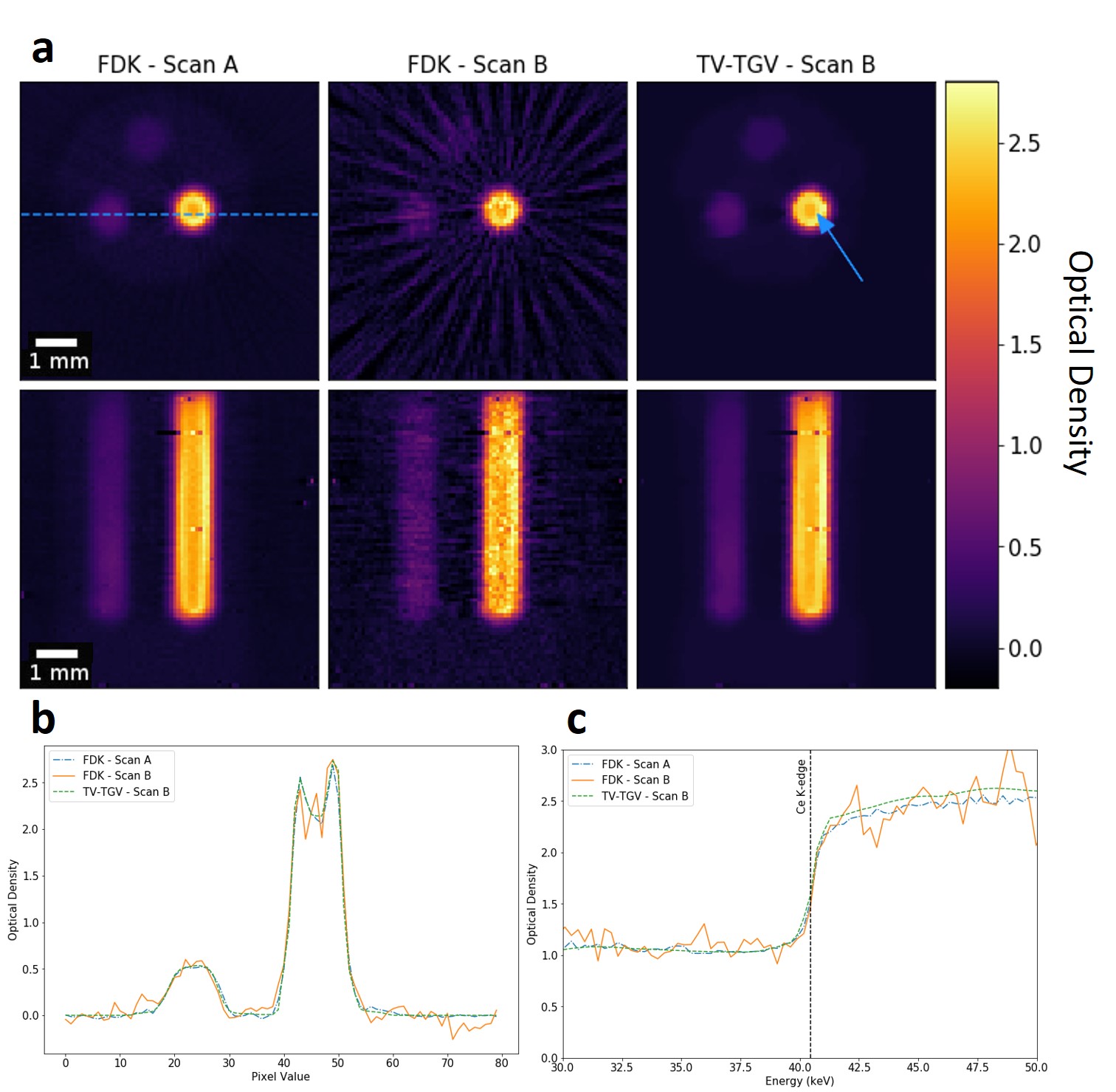}
	\caption{\textbf{Comparison of reconstruction algorithms.} \textbf{(a)} Transverse and frontal slices, showing reconstructions for FDK of sample Scan A (left column), followed by Scan B reconstructions with FDK (middle column), and TV-TGV (right column). Significant noise reduction is achieved for Scan B upon application of Total Variation in the spatial domain, comparable to feature definition achieved in Scan A. All reconstructed slices are shown for a single energy channel (42.27 keV). A dashed line and arrow indicate the positions from which spatial and spectral profiles were respectively measured for each reconstruction. \textbf{(b)} Spatial profile across two powder phases for the same energy channel. \textbf{(c)} Absorption spectra for a single voxel within the cerium powder region. All three reconstructions present a distinct absorption step, however improved linearity is obtained either side of the edge for Scan B upon application of TGV smoothing.}
	\label{fig:FDK_PDHG_full_comp}
\end{figure}

Figure \ref{fig:FDK_PDHG_full_comp}a shows a comparison of FDK and TV-TGV, through direct comparisons of reconstructed slices in the transverse and frontal planes. From the FDK reconstruction for Scan A, a distinct variation in OD is observed across the CeO$_{2}$ powder region, suggesting inhomogeneity in the powder. The frontal view confirms the non-uniform distribution, illustrating a consistently higher OD in the outer perimeter throughout the full depth of the sample. Both reconstructed views illustrate the sufficiently high count and sampling of the dataset, with minimal noise and strong feature edge definition. In contrast, the undersampled reconstructions of Scan B highlight the limitations of FDK for low numbers of projections, resulting in significant aliasing/streak artefacts, combined with increased noise across the reconstructed slices. 

Upon application of the TV-TGV method for Scan B, however, the data is recovered well, with the use of TV in the spatial domain providing enough smoothing to remove all streak artefacts, while maintaining edge preservation, restoring an image quality comparable to that of Scan A. The application of TV-TGV can be analysed further by comparing spatial and spectral profiles for each method, as shown in Figures \ref{fig:FDK_PDHG_full_comp}b and \ref{fig:FDK_PDHG_full_comp}c respectively. For the spatial profile, measured across two of the powder phases (Fig. \ref{fig:FDK_PDHG_full_comp}b), the advantages of the TV-TGV method for low count data are evident. Noise levels are significantly reduced for Scan B, compared to the equivalent FDK profile, with the TV-TGV method almost completely replicating the profile acquired for Scan A. Similarly for the spectral profile (Fig. \ref{fig:FDK_PDHG_full_comp}c), TGV regularisation improves the linearity of the regions either side of the cerium K-edge for Scan B. It should be noted that, for such a simple phantom, identification of the absorption edge is not an issue here, regardless of reconstruction algorithm. However, the results demonstrate the capability of handling noisy, few projection data, and achieving high image quality in fast scan acquisitions. In this case, a reduction in scan time by a factor of 36 has been shown.

\begin{figure}[t]
	\centering
	\includegraphics[scale=0.48]{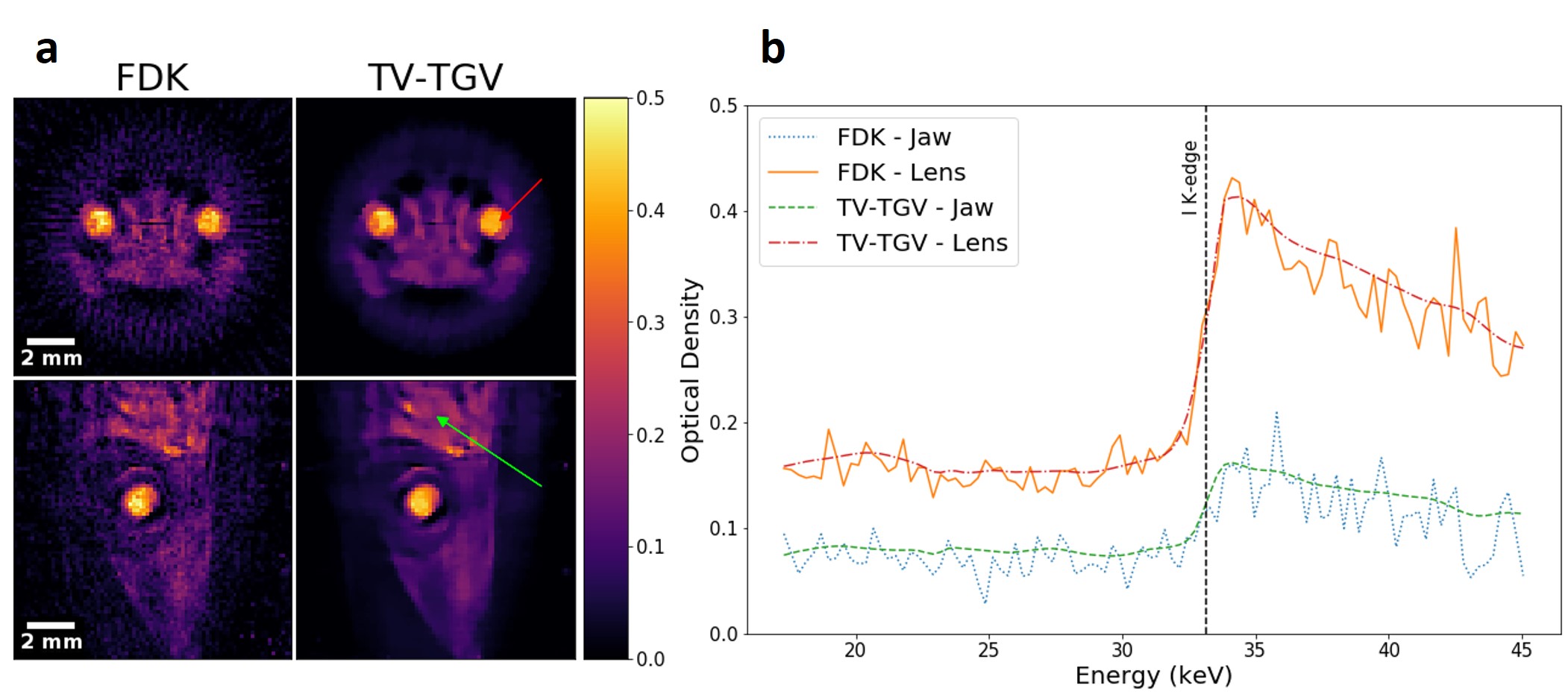}
	\caption{\textbf{Biological feature identification via regularised reconstruction.} \textbf{(a)} Reconstructed slices for channel 120 (33.95 keV), along both the axial and sagittal dimensions, following FDK (left column) and TV-TGV (right column) reconstruction. General noise reduction and smoothing due to TV regularisation is observed over all spatial regions. \textbf{(b)} Absorption spectra measured for a single voxel in two regions of the sample (red arrow - lens, green arrow - jaw). Significant smoothing due to TGV regularisation allows the presence of the iodine absorption edge to be more easily identified, in particular within the jaw adductor muscles of the specimen. A line signifying the theoretical position of the iodine K-edge is overlaid for comparison.}
	\label{fig:lizard_fdk_pdhg_spec}
\end{figure}

We next explore a more realistic sample in the field of biological imaging, where low SNR may pose problems in material identification and feature segmentation. An iodine-stained lizard head (Anolis sp.) was scanned at a beam voltage of 50 kV, at a maximum power of 0.7 W, reconstructed with a voxel size of 137 $\mu$m. A spectral subset of channels from 60-160 (energy range 17.3 - 45.0 keV) was chosen given the prior knowledge of the single iodine stain, eliminating the need to reconstruct the full spectral range for identification of key elemental signals. A reduced projection dataset was used, such that only 60 projections (at 120 s exposures) were reconstructed over the full 360$^{\circ}$ rotation, to test the capabilities of the reconstruction algorithms. The need to limit X-ray dose is more prevalent in the life sciences field and, as such, the ability to extract high quality information from low count and/or undersampled data is crucial. Once more, TV-TGV reconstruction was applied, with the results compared to the conventional FDK method. Here, initial parameter estimates were based on the optimal conditions achieved in the reconstruction of the phantom sample, allowing faster identification of the optimal parameters. Final reconstruction was performed with parameters of 0.002, 0.25 and 0.35 for $\alpha$, $\beta_{1}$ and $\beta_{2}$ respectively, with a total runtime of 40 minutes following 1000 iterations.
Figure \ref{fig:lizard_fdk_pdhg_spec} demonstrates the smoothing effects in both the spatial and spectral domains, with the lens and jaw adductor muscles highlighted as key regions of interest. For both the axial and sagittal reconstructed slices (Fig.  \ref{fig:lizard_fdk_pdhg_spec}a), spatial smoothing provided improved edge definition of the exterior head shape, as well as the outer eye and jaw regions. The presence of the iodine contrast agent is easily confirmed with the precise matching of the theoretical edge position (33.169 keV) in each spectral profile (Fig. \ref{fig:lizard_fdk_pdhg_spec}b). The benefit of noise suppression in the energy domain is demonstrated, allowing the presence of absorption edges to be more clearly defined, where they can often be lost within noisier reconstructions like FDK. Areas of lower iodine uptake, such as that of the jaw adductor muscles, may therefore be confidently identified as iodine-containing structures.

\begin{figure}[t]
	\centering
	\includegraphics[scale=0.56]{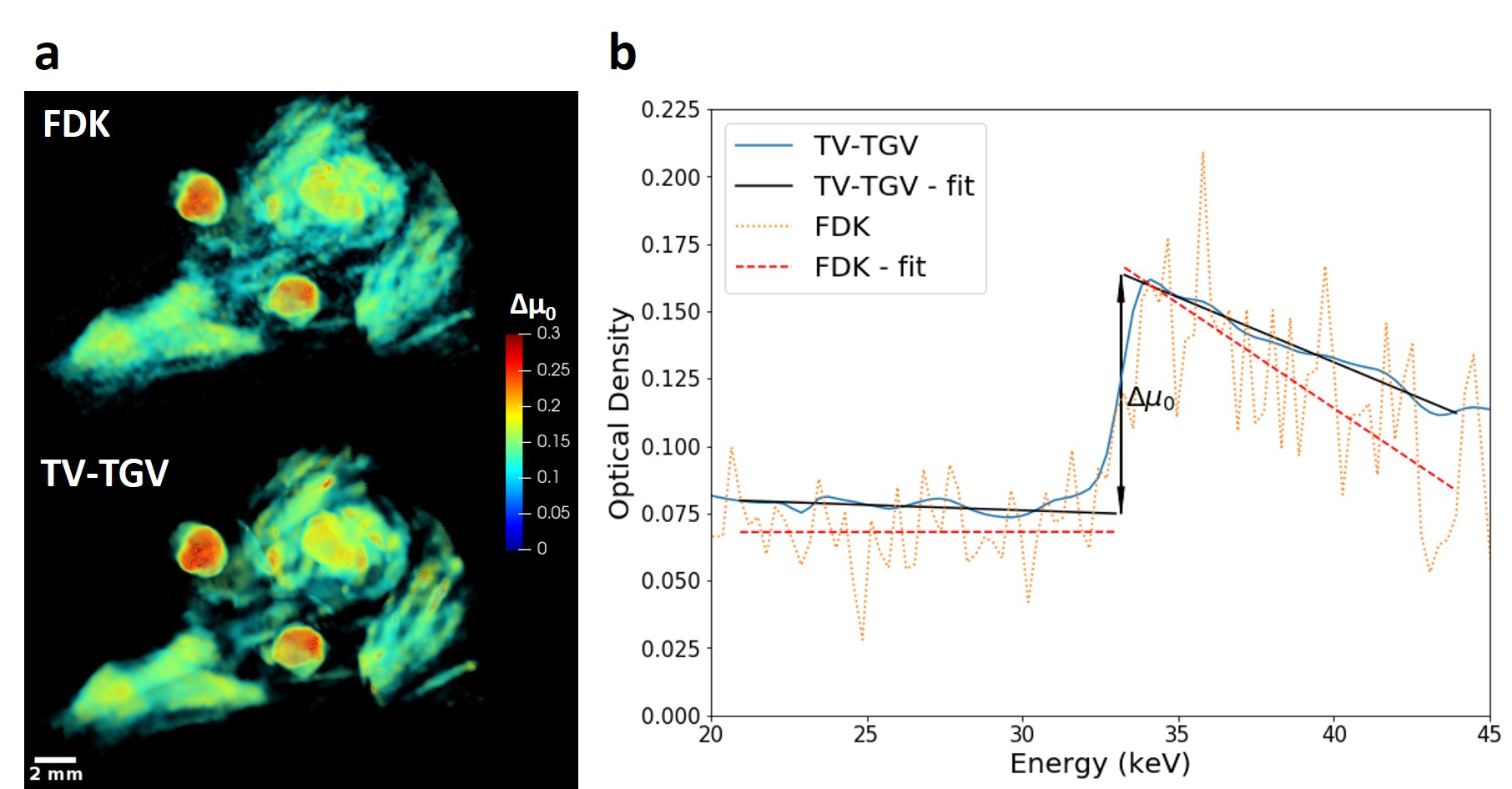}
	\caption{\textbf{Attenuation step size analysis for the iodine K-edge.} \textbf{(a)} 3D visualisations of the step size in the absorption edge, $\Delta\mu_{0}$, corresponding to relative iodine concentration. Images are shown for the lizard head sample following both FDK (upper left) and TV-TGV (lower left) reconstruction. \textbf{(b)} Absorption spectra acquired within the jaw adductor muscle for the same voxel in each reconstructed volume. Linear fits were acquired and extrapolated to the extremities of the absorption edge, where the relative change in optical density values was measured.}
	\label{fig:lizard_edge_height}
\end{figure}

Given the presence of an absorption edge, spectral analyses may be performed to provide information on iodine distribution throughout the biological specimen. Firstly we utilise the availability of a spectral profile in each voxel to measure relative iodine concentration, by virtue of spectral profile fitting. As shown in Fig. \ref{fig:lizard_edge_height}, linear least squares fitting was applied to regions before and after the absorption edge step for both the FDK and TV-TGV reconstructed volumes. By extrapolating and evaluating these fits at the known position of the K-edge (33.169 keV), we can precisely measure the size of the step change, $\Delta\mu_{0}$ \cite{Egan2015}. Repeating this process at every voxel provides us with a map of $\Delta\mu_{0}$, which is directly proportional to the concentration of iodine present in the sample. Calculated values of $\Delta\mu_{0}$ are shown for both reconstructed volumes in Fig.  \ref{fig:lizard_edge_height}a. Significant noise distortions in the FDK spectra lead to erroneous linear fitting, and consequently inaccurate measurements of $\Delta\mu_{0}$, as shown in Fig. \ref{fig:lizard_edge_height}b. Spectral smoothing due to TV-TGV, however, ensures improved precision in calculation of relative iodine concentration across the volume. Results indicate the diffusion of iodine fully into the lens, with high concentrations at the interior, and slightly lower levels on the exterior surface. Further, the increased reliability of $\Delta\mu_{0}$ measurements allows us to confidently identify 'hot spots' of higher iodine uptake, appearing on the brain and sections of the jaw muscles. The results are in good agreement with expected uptake regions of iodine contrast agent \cite{Metscher2009,Gignac2016}.

Our second analysis uses the absorption edge as a means of segmenting iodine-containing material from the remaining structures. For this, we used K-edge subtraction (KES). That is, spectral information is extracted from energy channels before the edge of interest, and subtracted from an equivalent set after the edge. The result is a dataset containing only the contrasting material, eliminating other structures where attenuation is slight across this energy range. The method of KES has previously been applied for monochromatic imaging either side of absorption edges for segmentation \cite{Panahifar2016,Kulpe2019}, as well as in hyperspectral cases, highlighting its potential for segmenting materials where more than one K-edge is present \cite{Egan2015}. A detailed description of the method is provided in supplementary information (see Fig. S1). Here, KES also offers an opportunity to evaluate reconstruction quality by direct comparison of tissue segmentation for both the FDK and TV-TGV methods. 

\begin{figure}[t]
	\centering
	\includegraphics[scale=0.57]{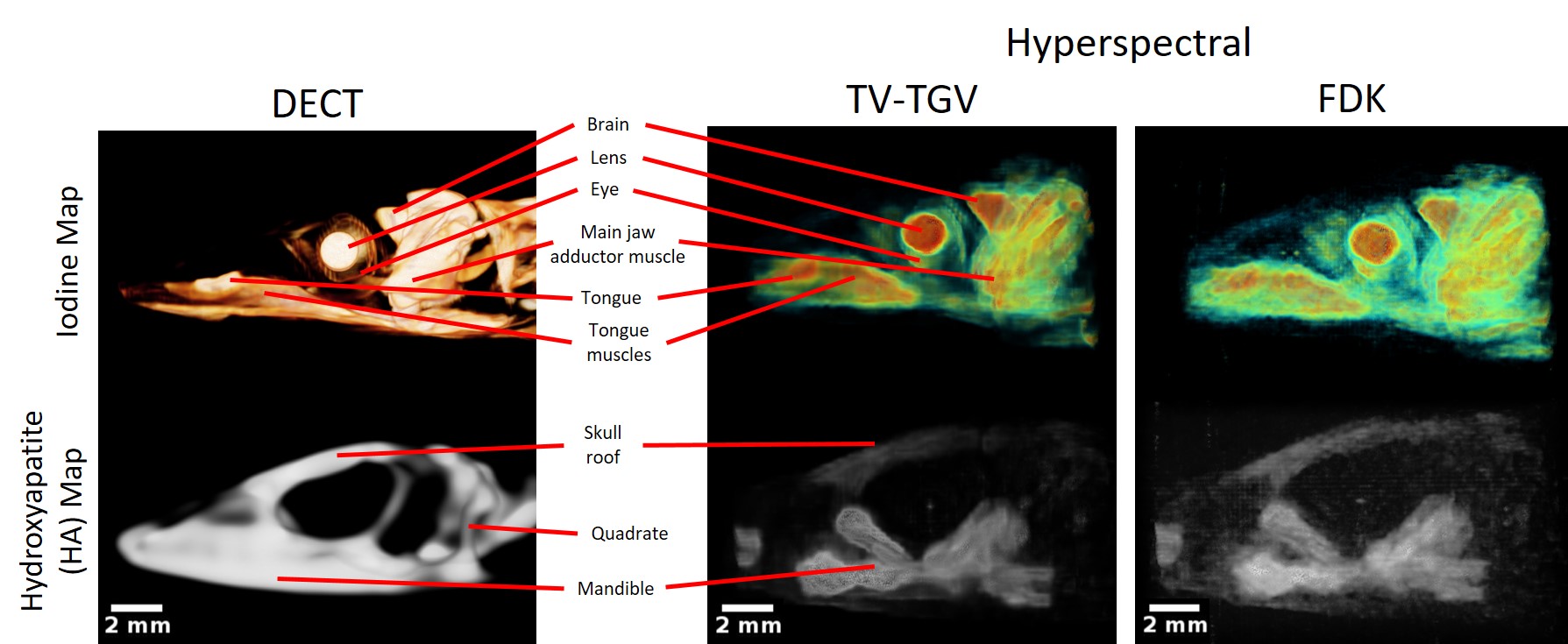}
	\caption{\textbf{Lizard head segmentation comparison for hyperspectral and dual-energy imaging.} Sagittal views of the segmented sample, producing maps of iodine-stained soft tissue (top row) and remaining hydroxyapatite (bottom row) bone structures. Results following K-edge subtraction for the TV-TGV reconstructed dataset (middle column) are directly compared to those following DECT acquisition of the same sample (left column), reduced to the same spatial resolution (137 $\mu$m). Labels indicate the successful segmentation of several iodine-stained soft tissue regions for the hyperspectral dataset, with similar structures identified in the DECT equivalent image. A comparison of HA maps show distinct bone structures observed across both datasets, as well as the accumulation of bone mineral in particular regions due to long term sample storage. Included is an example of a bone structure (quadrate) unidentified in TV-TGV segmentation. (Right column) Equivalent maps following FDK reconstruction of the hyperspectral dataset are also shown, with significant noise hiding a number of key features.}
	\label{fig:lizard_K_edge_sub}
\end{figure}

In order to measure the success of correct tissue segmentation, samples typically require a ground truth, or high quality dataset to precisely match the separation of distinct regions. For biological samples, however, this is not always available. Instead, we use a DECT scan acquired of the same sample, reduced to the same spatial resolution (137 $\mu$m) as that of the hyperspectral data. DECT has long been regarded as the 'gold standard' of biological stain imaging, and thus works well both as a measure of where hyperspectral X-ray CT stands in comparison, as well as an evaluation tool for each case of our spectral KES method. Segmented views of the sample are shown in Fig. \ref{fig:lizard_K_edge_sub} for both the TV-TGV regularised method, as well as the FDK reconstructed volume. The resulting 3D visualisations are shown upon hyperspectral KES around the iodine edge. Using the DECT segmentation as reference for identifying key soft tissue features, the advantages of TV-TGV over FDK become clear. The increased level of noise due to FDK leads to reduced visibility, particularly in regions of lower iodine concentration, such as the tongue and jaw adductor muscles. In contrast, following KES of the TV-TGV volume, clear separation is observed for regions to which the iodine has diffused. Structures including the brain, lens, tongue and jaw muscle all show strong X-ray signal enhancement due to sufficient staining by elemental iodine. Segmentation of the remaining material offers the ability to observe 'non-contrast-enhanced' structures. In this case, the external skull structure, consisting mostly of hydroxyapatite (HA), remains. As confirmed by the DECT results, visualisation of certain bone structures, including the skull roof and mandible region, are achieved. Full definition of the skull structure is lost however, and this is attributed to the long-term storage of the sample prior to imaging. As such, bone mineral has accumulated in some regions, while having dissipated in others. Therefore, precise segmentation of HA material was not expected. Nevertheless, the advantage of combined hyperspectral imaging and advanced reconstruction algorithms is provided through the successful segmentation of iodine in the biological structure. Moreover, while 60 projections with 120 s exposures were taken, results from our phantom sample suggest a further reduction in exposure time is possible with minimal loss in reconstruction quality, owing to the application of regularised algorithms. 

%% file: discussion.tex
\section*{Discussion} \label{sec:discussion}
The above case studies have highlighted the current state of the art for hyperspectral imaging in a lab-based setting and shown that, given an appropriate reconstruction algorithm, comparable levels of feature definition and characterisation of tissues and structures can be obtained to those in dual-energy and multispectral CT. However, our work is not without limitations. Spatial resolution is still far behind the standard set in conventional, and dual-energy, CT, emphasised by the downsampling of the aforementioned DECT results from their initial 9 $\mu$m voxel size. While currently limiting in the range of samples and features that may be observed, spatial resolution is only expected to improve with new detector technology. The use of an iodine contrast agent did not fully test the capability of chemical detection in hyperspectral imaging, given its high affinity for multiple soft tissue regions. A reasonable extension to the experiment is the use of additional chemical tracers, which bind to specific biological structures at varying concentrations. Given the high energy resolution of the hyperspectral (HEXITEC) detector, visualisation and segmentation of multiply-stained structures may be easily performed by virtue of measuring several spectral fingerprints. It has also previously been shown that absolute values of concentration may be determined in spectral imaging, given the use of an appropriate calibration phantom \cite{DeVries2015}, offering potential diagnostic insight on tracer concentration and distribution as a function of time. The use of an \emph{ex vivo} biological sample here may also be used as a step towards \emph{in vivo} structure analysis in the future. Biological scans performed \emph{in vivo} are limited more severely in terms of allowable X-ray dose and chemical concentration, hence the analysis performed in this work offers a solution in the form of regularised reconstruction for noisy, short exposure cases. Alternatively, there is a case to be made for the applicability of hyperspectral imaging in the absence of distinct spectral markers. With the availability of full spectral profiles at each pixel, beam-hardening artefacts may be eliminated now that changes in attenuation may be discriminated as a function of energy. In addition, while methods like K-edge subtraction are no longer applicable, measurement of relative attenuation changes over the energy range enables segmentation of poorly contrasting materials, which otherwise cannot be differentiated in conventional X-ray CT \cite{Egan2015}. Previous work on bone densitometry and soft tissue segmentation has been shown to be possible with few channel spectroscopic imaging, without the need for contrast agents \cite{Yokhana2017}. 

The use of advanced iterative reconstructions, in particular with application of regularisation terms, is also in its early stage development. The chosen method here of TV-TGV in the spatial-spectral domains is applicable given the composite materials producing sharp absorption edges, however optimisation of the regularised parameters for each dimension can be a slow process, due to the decoupled nature of the reconstruction algorithm. While every individual dataset will require specific tuning of each parameter, further development in the reconstruction protocols, combined with software packages like CIL to enable them, continues to allow these processes to be performed more quickly, while also laying a foundation upon which future methods may be based. In addition, the above studies have demonstrated the reduction in overall scan time enabled by the algorithms in CIL, offsetting the time taken to optimise reconstructed image quality. We predict that these methods will continue to allow further reductions in scan time, as well as limiting overall X-ray dose, without suffering significant losses in image quality, an advantage that may prove to be hugely beneficial to imaging in the biological field.

%% file: conclusion.tex
\section*{Conclusion} \label{sec:Conclusion}
This paper has applied hyperspectral imaging to a simple test phantom, as well as, for the first time, mapping the location of staining in a biological sample, using as many as two hundred energy channels (800 eV resolution). We have highlighted the vulnerabilities of conventional reconstruction methods for lab-based hyperspectral X-ray imaging. Through the use of a novel, spatiospectral reconstruction algorithm, we have enabled precise chemical identification and mapping at the micrometer scale. Examination of a multi-phase phantom emphasised the significant reductions (36 times shorter) in scan time achievable by implementing regularised reconstruction to compensate for noisy datasets. In performing an \emph{ex vivo} spectral CT scan of an iodine-stained lizard head sample, we have shown the capability of hyperspectral CT to have the elemental sensitivity to compete with existing techniques, such as DECT, in soft tissue segmentation and structural analysis, but with definitive identification of the iodine location through its characteristic K-edge. While here a single stain was measured and visualised, the exploration of multi-labelled biological samples is possible, given the high spectral resolution of the detector. Further, the weaknesses of analytic regimes such as FDK have been highlighted for spectral imaging, particularly for short scan acquisitions, reinforcing the need for standardised, iterative algorithms such as those provided in CIL. Together with the reduced scan times they enable, the correlated reconstruction methods open up the potential for hyperspectral studies in fields including non-destructive testing, security scanning and chemical catalysis. With improving detector technology and multi-staining methods, we conclude that lab-based hyperspectral CT offers great future prospects for biological research, among a number of other fields, such as chemical engineering, geology, materials science and cultural heritage.

%% file: methods.tex
\section*{Materials and Methods} \label{sec:methods}
\subsection*{Phantom Sample Preparation}
An aluminium cylinder was used as the matrix for three internal powders, due to its low attenuation relative to other metals. The external matrix had measured dimensions of 0.5 cm in diameter, and 1 cm in height. Three cylindrical holes (diameter 0.7 mm) were drilled to a sample depth of 75\% (0.75 cm), and filled with a different powder (CeO$_{2}$, ZnO and Fe).

\subsection*{Biological Sample Preparation}
For this experiment, the head of a lizard (Anolis sp.) was scanned and analysed. For long-term stability, the sample was fixed in formalin and stored in 70\% ethanol, prior to staining. The sample was then dehydrated to 100\% ethanol, before staining with 1\% elemental iodine in absolute ethanol (I$_{2}$E). It has previously been shown that I$_{2}$E offers strong contrast in soft tissue, allowing for discrimination from bone and teeth (hydroxyapatite) structures \cite{Metscher2009,Handschuh2017}. After staining, the sample was washed with 100\% ethanol. Finally the biological specimen was mounted in 1.5\% Agarose.

\subsection*{X-ray Detector}
The hyperspectral imaging was performed using an energy-sensitive HEXITEC detector \cite{Seller2011}, consisting of a 1 mm thick CdTe single-crystal semiconductor, bump-bonded to an ASIC producing a 2 cm $\times$ 2 cm detection area. The system is split into an 80 $\times$ 80 pixel array, with a 250 $\mu$m pitch. The detector offers an energy resolution of approximately 800 eV at 59.5 keV and 1.5 keV at 141 keV. All raw data was acquired on an event-by-event basis using the HEXITEC detector software. 

\subsection*{Data Acquisition Routines}
For imaging of the phantom sample, a parallel source-sample-detector configuration was implemented in the custom-built Colour Bay, part of the Henry Moseley X-ray Imaging Facility (HMXIF) at The University of Manchester. The walk-in X-ray bay contains a 225 kV source, and full manipulator control is available via MATLAB scripts. Two scans of the phantom were acquired. Identical scanning conditions were implemented, at a geometrical magnification of 2.54, with the polychromatic X-ray source operating at a tube voltage of 60 kV with beam power 6 W. This was such that contrast was optimised, while remaining within the count-rate limit of the detector. Exposure time, however, was varied, from 30 s up to 180 s per projection. In total, 180 projections were acquired per scan, with 2$^{\circ}$ step size over a full 360$^{\circ}$ rotation, resulting in total scan times of 2.5 and 11 hours, accounting for buffer times and bias voltage refreshing between projections in single photon detection \cite{Seller2011}. Prior to reconstruction, one dataset was downsampled to 30 projections with 30 s exposure time to evaluate reconstruction quality on undersampled data.

As a further demonstration of the flexibility in lab-based hyperspectral imaging, the HEXITEC was next combined with the Nikon High Flux Bay system at the HMXIF. Once more a 225 kV source was utilised, however improved contrast was possible due to lower flux capabilities, necessary for imaging of soft tissue within the biological sample. Source and sample manipulation were controlled by Nikon's proprietary software Inspect-X. A software module was created using the IPC interface to Inspect-X \cite{Gajjar2017,Gajjar2018}, enabling communication between Nikon software and the spectral detector. The biological sample was secured to the rotation stage, held in place such that the specimen was suspended vertically during the acquisition. A geometric magnification of 1.81 was determined to project the sample fully in the detection area. The polychromatic X-ray beam was operated at a peak voltage of 50 kV, at a maximum power of 0.7 W. Projection images were recorded at an angular step size of 2$^{\circ}$ over a full rotation, with exposure times of 120 s for each of the 180 projections, for an 8 hour total scan time. A reduced subset of the dataset was later taken for reconstruction, as described in the main text.
For all scans throughout the study, sets of four flat-field projections were acquired both before and after scanning for fixed-pattern noise subtraction, while a further dark current correction was applied during set-up. Detected events were binned into a set of spectral channels, with the number determined by the maximum X-ray energy.

Prior to scanning, a preliminary energy calibration procedure was performed through the measurement of characteristic fluorescence signals from a series of metals \cite{Alkhateeb2013}. The calibration procedure provides a direct transition between energy channels and their corresponding energies. Further, an inter-pixel gain correction was applied through the use of a correlative optimised warping algorithm using the same data \cite{Tomasi2004,Egan2017}. 

The accompanying DECT scan was conducted using a Zeiss Xradia XRM-400. The dual-energy acquisition performed scans at 40 kV and 80 kV, with a 0.17 mm glass filter and 0.4x optical lens. A total of 1051 projections were acquired over 210$^{\circ}$ rotation, with 15 s exposure times. The final reconstructed volume had a voxel size of 9.4 $\mu$m.

\subsection*{Spectral CT Data Reconstruction}

Initial processing of the acquired datasets was handled via MATLAB routines. In the case of the biological sample, a combined wavelet-based Fourier filter was applied in every channel for the removal of ring artefacts across the dataset \cite{Munch2009}. For each dataset, the resulting 4D matrix (3 spatial dimensions, 1 spectral dimension) was then reconstructed using the CIL software. Full functionality and operation of the Python-based framework has been discussed elsewhere \cite{Jorgensen2020,Papoutsellis2020}. Given its traditional use in conventional cone-beam reconstruction, the FDK method formed the baseline from which all other results were compared and evaluated. For comparison, an iterative reconstruction algorithm was chosen, combining data fitting with spatial and spectral regularisation terms. The iterative algorithm took the form:
\begin{equation} \label{eq:it_recon}
\underset{u}{\text{min}} ||Au-b||^{2}_{2} + \alpha \textrm{TV}_{x,y,z}(u) + \beta_{1,2} \textrm{TGV}_{c}(u)
\end{equation}
where our first term concerns classic least-squares data fitting, with hyperspectral projection data, $b$, related to $u$ through the operator, $A$, based on the system geometry and properties. The latter terms concern the addition of spatial and spectral regularisation. Since 4D datasets exhibit different image properties in the spatial and spectral dimensions, two different regularisation terms were applied for their noise reduction capabilities. The result is a 'decoupled' regularisation algorithm. We define $\textrm{TV}_{x,y,z}(u)$ as the Total Variation (TV), applied across each spatial dimension \cite{Rudin1992} for a single channel, before summing over all channels (channel-wise). TV is one of the most widely used regularisers, as it favours piece-wise constant images, with sharp edge boundaries. The application of TV therefore allows for noise suppression of flat signal regions, while maintaining the local discontinuities at structure edges. The TV model fits well for CT images and has been successfully used previously for undersampled CT data \cite{Sidky2006}, as well as noise suppression purposes \cite{Rudin1992,Chan2005}. The use of TV regularisation has also previously demonstrated its benefits in spectral image reconstruction over analytic methods like FDK \cite{Xu2014,Rigie2015}. However, TV is known for introducing ’staircasing’ artefacts for piece-wise affine or smooth signals, for example ramp structures, which can result in patchy, unnatural reconstructed images \cite{Chan2005}. Further, as the aim of TV regularisation is to reduce unwanted signal variations such as noise, the final reconstruction can suffer from a loss of contrast due to reductions in intensity \cite{Chan2005}. For the spectral domain, we define $\textrm{TGV}_{c}(u)$ as the Total Generalised Variation (TGV), based on the method proposed by Bredies et al. \cite{Bredies2010}. Here we apply TGV regularisation along the channel direction, for each individual voxel in the image. TGV becomes applicable in the case of K-edge imaging, where we expect a smoothed step shape to our energy profiles across the edge position, spanning a few channels. TGV is able to reduce noise variations while maintaining the definition of any absorption edges present. Further, TGV avoids the staircasing effects experienced with TV. CIL provides a number of iterative algorithms to solve Equation \ref{eq:it_recon}, here we applied the primal dual hybrid gradient (PDHG) method \cite{Chambolle2011}. Three regularisation parameters ($\alpha$ for TV, $\beta_{1,2}$ for TGV) were used to control the strength of penalisation for each regulariser, and muse be optimised for each dataset. Therefore parameter values were chosen to suppress noise without significant blurring of reconstructed images or spectral profiles. In the case of noise suppression, optimal parameters vary based on the noise level in the dataset. We direct the reader elsewhere for further details on the different approaches adopted in previous studies for defining parameter ranges \cite{Knoll2011,Holler2014,DelosReyes2017}. For the phantom sample, optimal values were determined based on three factors: visual comparison with reconstructed slices of Scan A, precision of absorption edge position, and minimum obtained 'primal-dual gap'. The latter is a quantitative measure of convergence for the PDHG method, and has previously been used as a stopping criterion for the iterative method \cite{Zhu2008,Chambolle2011}.